\documentclass[aip,apl,twocolumn,amsmath,amssymb,floatfix,reprint,square,comma,numbers]{revtex4-1}
\usepackage{graphicx}
\usepackage{dcolumn}
\usepackage{bm}
\usepackage[colorlinks,linkcolor=red, anchorcolor=blue,citecolor=blue]{hyperref}
\usepackage{natbib}

\begin{document}
\title{Synthesizing three-body interaction of spin chirality with superconducting qubits}

\author{Wuxin Liu}
\thanks{W. L. and W. F. contributed equally to this work.}
\author{Wei Feng}
\thanks{W. L. and W. F. contributed equally to this work.}
\author{Wenhui Ren}
\author{Da-Wei Wang}
\email{dwwang@zju.edu.cn}
\author{H. Wang}
\email{hhwang@zju.edu.cn}
\affiliation{$^{1}$ Interdisciplinary Center for Quantum Information and \mbox{Zhejiang Province Key Laboratory of Quantum Technology and Device,}
\mbox{Department of Physics, Zhejiang University, Hangzhou 310027, China}}

\date{\today}
\begin{abstract}
Superconducting qubits provide a competitive platform for quantum simulation of complex dynamics that lies
at the heart of quantum many-body systems, because of the flexibility and scalability afforded by the nature of microfabrication.
However, in a multiqubit device, the physical form of couplings between qubits is either an electric (capacitor) or magnetic field (inductor),
and the associated quadratic field energy determines that only two-body interaction in the Hamiltonian can be directly realized.
Here we propose and experimentally synthesize the three-body spin-chirality interaction in a superconducting circuit based on Floquet engineering.
By periodically modulating the resonant frequencies of the qubits connected with each other via capacitors,
we can dynamically turn on and off qubit-qubit couplings, and
further create chiral flows of the excitations in the three-qubit circular loop. 
Our result is a step toward engineering dynamical
and many-body interactions in multiqubit superconducting devices, which potentially expands the degree of freedom
in quantum simulation tasks.
\end{abstract}
\maketitle
Quantum simulation is an efficient way to solve the classically inaccessible
many-body problems of complex quantum systems \cite{Feynman1982}.
Among different strategies, analog quantum simulation, which intuitively
uses a controllable quantum system to mimic the Hamiltonian of a target
one, is more attainable in the near future because of its higher tolerance level of errors \cite{Nori2014RMP,Somaroo1999}.
Superconducting qubits with advantages in flexibility and scalability are
one of the leading candidates for building a practical analog quantum simulator~\cite{Clarke2008,You2011,Devoret2013,Martinis2018Science}.
In a multiqubit superconducting circuit, qubits are typically connected by capacitors/inductors
which store electric/magnetic fields, and the coupling energy contribution to
the Hamiltonian depends quadratically on the field operators.
Therefore it is straightforward to implement the two-body flip-flop interaction in the isotropic spin-1/2
$XY$-type model through the capacitive or inductive coupling \cite{wendin2017}. However, interactions containing
more spin operators, which are useful in calculating the ground state
energy of molecules \cite{VQEprx,Kandala2017} and in toric codes \cite{Kitaev2003},
remain challenging to be experimentally realized.

Floquet engineering characterized by fast periodic modulation of the characteristic frequencies of a quantum system is a versatile
way to control the long-time dynamics of the system \cite{Goldman2014PRX}. The
key of Floquet engineering is designing appropriate driving schemes
to synthesize the target Hamiltonian. In superconducting qubits, Floquet engineering
has been used to realize chiral ground state current\cite{Martinis2017natphy},
quantum switch\cite{zhuxiaobo2018switch}, perfect state transfer\cite{sunluyan2018statetransfer},
Dzyaloshinskii-Moriya (DM) interaction\cite{dawei2019natphy}, and
topological magnon insulator states\cite{SunLuyan2019}. Although it has been
proven in theory that arbitrary two-body spin interactions can be
synthesized by Floquet modulation\cite{Floquet2018,Floquet2019},
many exotic phenomena, such as topological phases with anyonic excitations \cite{wenxiaogangBOOK},
can only appear as a direct consequence of the many-body interactions involving more than two spins \cite{Hafezi2014prb}.

Spin chirality \cite{wenxiaogang1989} represents one example of the many-body interactions involving three spins,
$\hat{\chi}=\bm{\sigma_{1}}\cdot\left(\bm{\sigma_{2}}\times\bm{\sigma_{3}}\right)$, 
where $\boldsymbol{\sigma}_{j}=\left(\sigma_{j}^{x},\sigma_{j}^{y},\sigma_{j}^{z}\right)$
is the Pauli vector for the $j$th spin-1/2 particle. It
plays an important role in quantum Hall effect in magnetic
materials\cite{Taguchi2001,Grohol2005,Bruno2004,Katsura2010,Hirschberger2015,Lyanda2001}
and chiral spin states in spin liquids\cite{Bauer2014,zhou2017}.
The dynamics driven by $\hat{\chi}$ features a chiral evolution of
the three-spin states $|s_{1}s_{2}s_{3}\rangle$, i.e.,
$e^{-i\hat{\chi}\theta}|s_{1}s_{2}s_{3}\rangle=|s_{3}s_{1}s_{2}\rangle$, 
with $\theta=\pi/3\sqrt{3}$ and $s_{j}=0$ (spin down) or $1$ (spin up).
It can be used in perfect state transfer\cite{Christandl2004}
for arbitrary three-spin states. By changing $\hat{\chi}\rightarrow-\hat{\chi}$,
the rotation direction of the spin states is reversed. It is worth
noting that $\hat{\chi}$ breaks both the time reversal symmetry $T$
(replacing $\boldsymbol{\sigma}_{j}$ with $-\boldsymbol{\sigma}_{j}$)
and parity symmetry $P$ (exchanging $\boldsymbol{\sigma}_{i}\leftrightarrow\boldsymbol{\sigma}_{j}$)
but conserves the $PT$ symmetry\cite{wenxiaogang1989}.

Here we propose and experimentally demonstrate a method to synthesize the three-body spin chirality interaction $\hat{\chi}$,
based on Floquet engineering for three superconducting qubits that are pair-wisely connected.
By periodically modulating the resonant frequencies
of the three qubits with well-controlled amplitudes, periods, and phases, we can dynamically
turn on and off the coupling between any two qubits, and further create the three-body term $\hat{\chi}$ in
the effective Hamiltonian. Under the engineered $\hat{\chi}$, we characterize the three-qubit
chiral dynamics where excitations injected to one or two of the qubits flow
clockwise through the three-qubit sites, which is in excellent agreement with numerical simulation taking into account
qubit anharmonicity.
The capability of engineering dynamical
and many-body interactions beyond the conventional two-body ones will certainly expand the functionality
of the quantum simulation platform built upon multiqubit superconducting circuits.

\begin{figure}[t]
\includegraphics[width=1\linewidth]{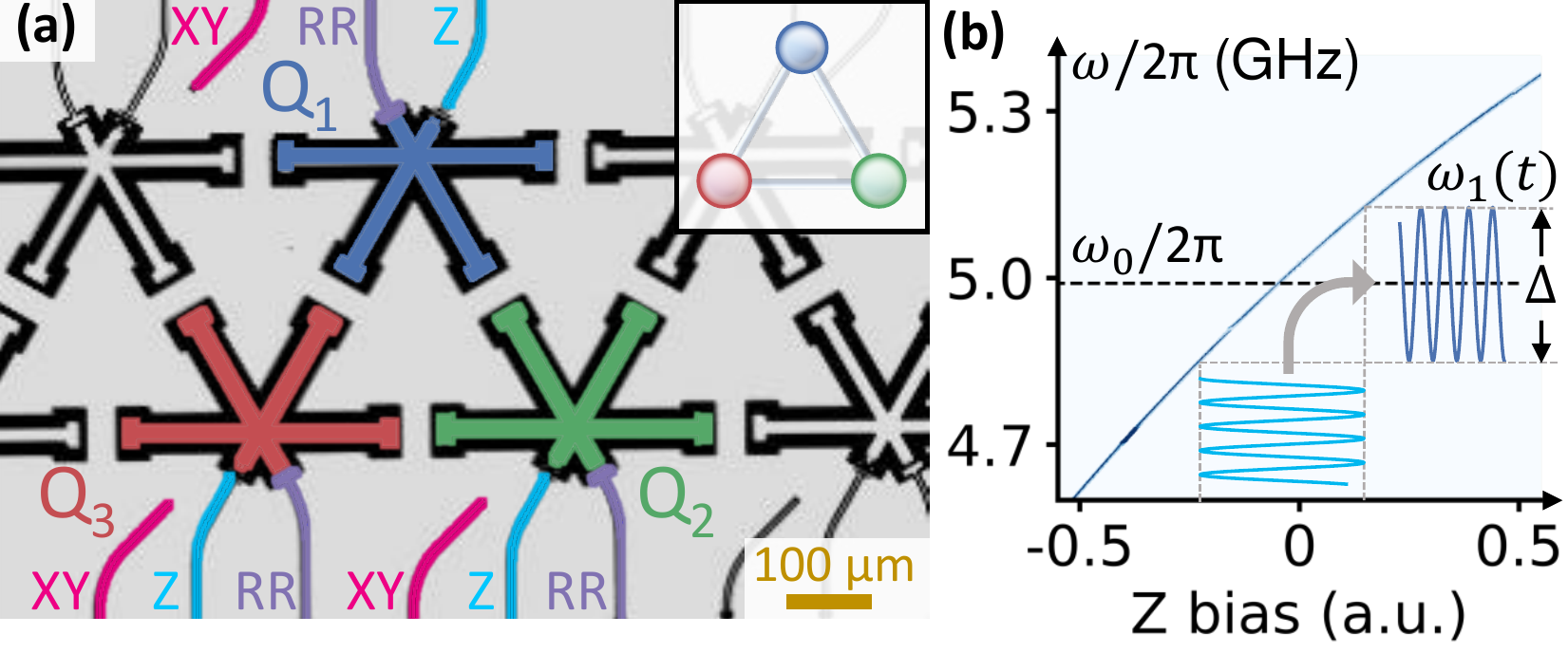} \caption{\label{fig:sample}
(a) False color image of the three qubits, as labeled, for synthesizing the three-body interaction $\hat{\chi}$.
Inset: Cartoon illustration of the three qubits arranged in a circular loop.
(b) Resonant frequency as a function of Z line bias for $Q_1$, which
is used to calibrate the nearly-sinusoidal Z pulse (light blue sinusoid)
for modulating the qubit frequency with the desired amplitude $\Delta$ (dark blue sinusoid) .
}
\end{figure}

The multiqubit superconducting device used in this experiment is illustrated in Fig.~\ref{fig:sample}(a), where
frequency-tunable transmon qubits \cite{Xmon,Song2019} are arranged in a triangular chain, and each qubit is
capacitively coupled to four neighbors with the coupling strength
$g/2\pi$ ranging from {10 to 13}~MHz. To synthesize $\hat{\chi}$, 
we focus on the three qubits $Q_1$, $Q_2$, and $Q_3$ as highlighted by colors of blue, green, and red in Fig.~\ref{fig:sample}(a), respectively, while
all the other qubits (light gray) are far detuned in frequency.
Assuming that the pair-wise coupling strengths are uniform, the three-qubit Hamiltonian is
\begin{equation}
H=\hbar\sum_{j=1}^{3}\omega_{j}(t)\sigma_{j}^{+}\sigma_{j}^{-}+\hbar g \sum_{jk}(\sigma_{j}^{+}\sigma_{k}^{-}+\sigma_{j}^{-}\sigma_{k}^{+}),
\label{hinit}
\end{equation}
where $\sigma_{j}^{+}$ and $\sigma_{j}^{-}$ are $Q_j$'s raising and lowering operators, respectively.
As labeled in Fig.~\ref{fig:sample}, each qubit $Q_j$ has a microwave drive line (magenta) to inject excitations for XY rotations,
a flux line (cyan) to tune the resonant frequency $\omega_{j}(t)$ for Z rotations, and a readout resonator (purple) for
detecting the state of the qubit. 
Single-qubit $\pi$ rotations used in the experiment are calibrated to be around {0.997} in fidelity by randomized benchmarking. Detailed device parameters such as qubit coherence times, readout fidelities, and qubit-qubit coupling strengths can be found in Supplementary Material.

As illustrated in Fig.~\ref{fig:sample}(b), with Floquet engineering, we apply periodic driving signals through
the Z line of $Q_j$, so that $Q_j$'s resonant frequency varies over time
according to $\omega_{j}(t)=\omega_{j}(0)+\Delta_j\cos(\nu_j t+\phi_{j})$, where values of
$\omega_{j}(0)$, $\Delta_j$, $\nu_j$, and $\phi_j$ are properly chosen
depending on the function to be achieved ($\Delta_j$, $\nu_j \gg g$ ). The Floquet driving signals can be
applied to one, two, or all three qubits simultaneously, and for simplicity here
we assume $\omega_{j}(0) \equiv \omega_0$, $\Delta_j \equiv \Delta$, and $\nu_j \equiv \nu$ for all qubits being modulated.
For example, in the interaction picture,
Eq.~(\ref{hinit}) becomes
\begin{equation}
H_I = \hbar g \sigma_{j}^{+}\sigma_{k}^{-} e^{i(\Delta/\nu)\left[\sin\left(\nu t+\phi_j\right)-\sin\left(\nu t+\phi_k\right)\right]} + \textrm{h.c.},
\label{EI}
\end{equation}
if both $Q_j$ and $Q_k$ are being modulated while the third qubit is far detuned and can be ignored.
Using the Jacobi-Anger expansion $e^{iz\sin(y)}=\sum_{n=-\infty}^{+\infty}J_{n}(z)e^{iny}$,
where $J_n(z)$ is the $n$th-order Bessel function of the first kind, the interaction Hamiltonian can
be expanded to
\begin{equation}
H_{I}=\sum_{n=-\infty}^{+\infty}H_{n}e^{in\nu t},
\label{EN}
\end{equation}
where $H_n$ contains the two-body operators $\sigma_{j}^{+}\sigma_{k}^{-}$.

\begin{figure}[b]
\centering{}\includegraphics[width=1\linewidth]{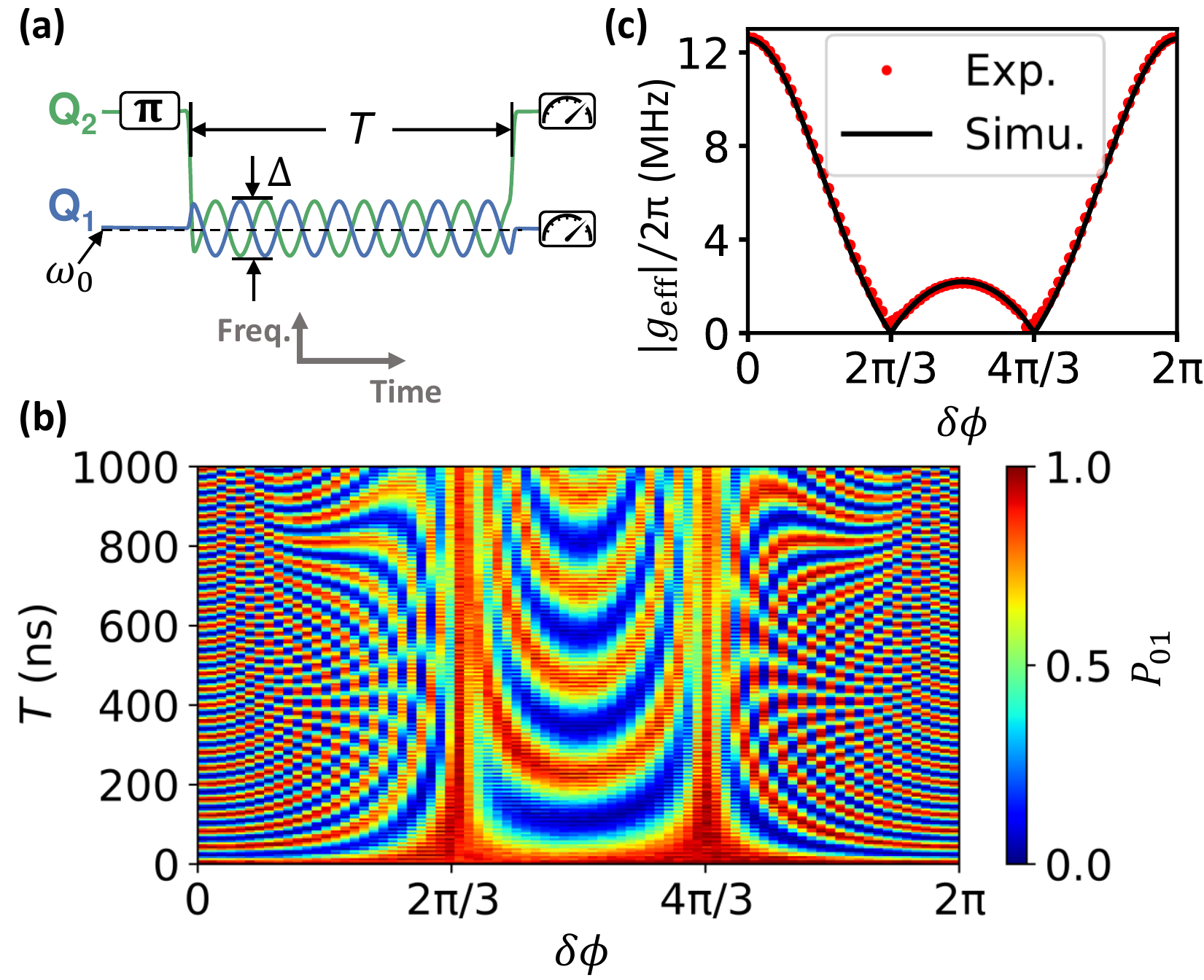}
\caption{(a) Pulse sequence to tune the effective coupling strength $g_\text{eff}$ between
$Q_1$ and $Q_2$ by applying Floquet driving signals to the two qubits simultaneously.
The sequence starts with initializing both $Q_1$ and $Q_2$ in $|0\rangle$ ($Q_3$ is far detuned),
and then exciting $Q_{2}$ to $|1\rangle$ with a $\pi$ rotation,
following which both qubits are modulated via sinusoidal signals centering
around {4.990 GHz} through their Z lines for a period of $T$.
The two sinusoidal signals have the same amplitude $\Delta$ and frequency $\nu$, but differ
in phase by $\delta\phi$.
Finally the two-qubit joint state is measured after the Floquet modulation.
(b) Experimental probability of the $Q_1$-$Q_2$ joint state ${|01\rangle}$ as a function of $T$ and $\delta\phi$.
At fixed $\delta \phi$ along the time axis, Rabi oscillations with varying periods are clearly observed.
(c) Effective coupling strength {$|g_\text{eff}|$} obtained by Fourier transform of the data in (b) as a function of $\delta\phi$ (dots),
in comparison with the numerical simulation result (line). \label{fig:swap}}
\end{figure}

For two qubits, a direct consequence of Eq.~(\ref{EN}) is that the time-independent zeroth-order term in $H_I$
dominates the long-time dynamics ($\gg 2\pi/\nu$) while the net contribution of all higher order terms ($|n| \ge 1$) is zero.
The zeroth-order term determines that the effective coupling strength, $g_\textrm{eff}$, between $Q_j$ and $Q_k$ under the Floquet engineering
is $gJ_{0}\left(2\sin(\delta\phi/2)\Delta/\nu\right)$, which can be tuned by varying
the ratio of $\Delta/\nu$ and the relative phase between the two Floquet driving signals $\delta\phi = \phi_{j}-\phi_{k}$.
An experiment is done to verify such a tunability based on the pulse sequence illustrated in Fig.~\ref{fig:swap}(a),
in which Rabi oscillations between $Q_1$ and $Q_2$ are mapped out at different $\delta\phi$ values
with $\omega_{0}/2\pi{\approx 4.990}\ \mathrm{GHz}$, $\nu/2\pi=100\ \mathrm{MHz}$, and $\Delta/2\pi{\approx 138}\ \mathrm{MHz}$.
At $\delta\phi=2\pi/3$ or $4\pi/3$, the oscillation patterns disappear as $g_\textrm{eff} = 0$,
which indicates that the two qubits are dynamically decoupled.
In Fig.~\ref{fig:swap}(c), ${|g_\textrm{eff}|}$ as a function of $\delta\phi$ (dots)
is obtained by Fourier transform of the data in Fig.~\ref{fig:swap}(b)
along the time axis, which is excellent agreement with the numerical simulation result (line).
We note that the tunability of $g_\textrm{eff}$ can also be realized by modulating just one of the two qubits. However, the modulation
amplitude $\Delta$ has to be about twice larger in order to completely decouple the two qubits,
which is experimentally more difficult and may further expose the qubit to the impact of two-level state defects in the spectrum (see Supplementary Material).

Now we show that by applying periodic driving signals through
the Z lines to all three qubits simultaneously, so that $Q_j$'s resonant frequency
is modulated according to $\omega_{j}(t)=\omega_{0}+\Delta\cos(\nu t+2\pi j/3)$, it is possible to
observe the long-time dynamics governed by the three-body $\hat{\chi}$ interaction.
The strategy is the same as before: Take the summation over all pair-wise terms similar to that in Eq.~(\ref{EI}) and
follow the Jacobi-Anger expansion to obtain Eq.~(\ref{EN}) for the three-qubit case,
where 
$H_{n}=\hbar gi^{n}J_{n}\left(\sqrt{3}\Delta/\nu\right)e^{i n(j+k)\pi/3}\sum_{jk}\left[\sigma_{j}^{+}\sigma_{k}^{-}+\left(-1\right)^{n}\sigma_{j}^{-}\sigma_{k}^{+}\right]$.
Note that $H_{n}$ is the Hermitian conjugate of $H_{-n}$.
At $\nu\gg g$, the second order perturbation is valid, and the effective
Hamiltonian governing the long-time dynamics is~\cite{Goldman2014PRX,dawei2016}
\begin{equation}
H_{\textrm{eff}} \approx H_{0}+\sum_{n=1}^{\infty}\frac{1}{n\hbar\nu}\left[H_{n},H_{-n}\right],
\label{Heff}
\end{equation}
where
$\left[H_{n},H_{-n}\right] \propto \sum_{jkl}\sigma_{j}^{z}\left(\sigma_{k}^{x}\sigma_{l}^{y}-\sigma_{k}^{y}\sigma_{l}^{x}\right)$.
Finally, with the chirality operator $\hat{\chi}$, we rewrite 
\begin{equation} 
H_{\textrm{eff}}=H_{0}+\hbar\kappa\hat{\chi},
\label{Heff1}
\end{equation}
where $\kappa=g^{2}\beta/\nu$ and $\beta=\sum_{n=1}^{\infty}J_{n}^{2}(\sqrt{3}\Delta/\nu)\sin(n\pi/3)/n$ (see Supplementary Material).

\begin{figure}[tb]
\centering{}\includegraphics[width=1\linewidth]{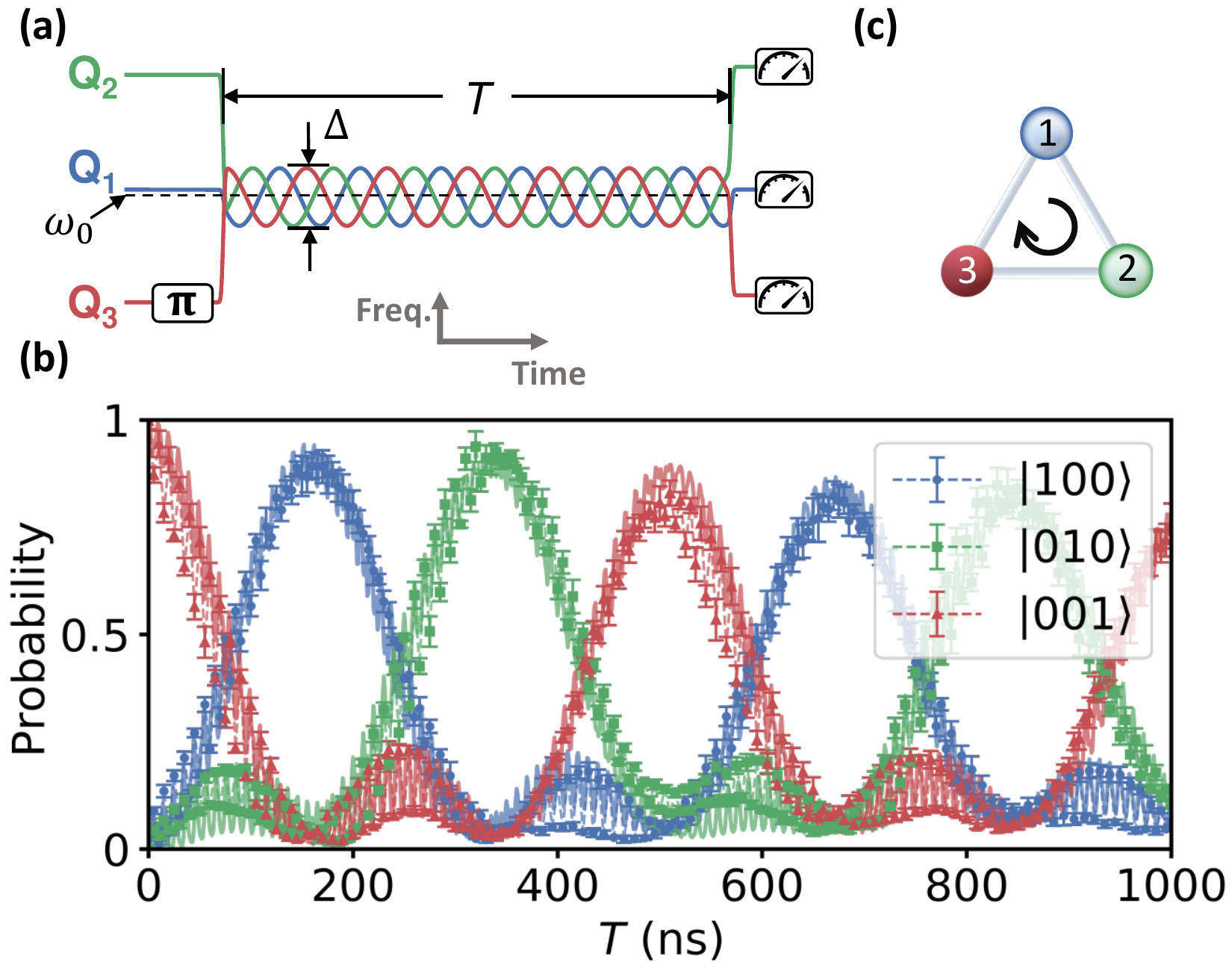} \caption{\label{fig:expdata_single}
(a) Pulse sequence for observing the single-excitation chiral dynamics.
(b) Measured probabilities of the single-excitation multiqubit states (symbols connected by dashed lines)
as functions of the modulation time $T$. Data curves of $P_{100}$, $P_{010}$, and $P_{001}$, 
colored according the site colors containing the excitation in (c), rise up
sequentially and cyclically, indicating the chiral motion of the single excitation. Error
bars are standard deviations of five repetitive measurements. Solid lines are the numerical simulation results.
(c) Schematics of the chiral dynamics, where the dark (light) color
represents the qubit in $|1\rangle$ ($|0\rangle$) and the arrow indicates the rotation direction of the single excitation.}
\end{figure}

Under the engineered $\hat{\chi}$, excitations injected to one or two of the qubits will flow
clockwise through the three-qubit sites (see Supplementary Material for the tune-up procedure). 
Figure~\ref{fig:expdata_single} shows such an
experiment where a single excitation injected to one of the qubits travels cyclically through the
three-qubit sites under the Floquet modulation.
The modulation parameters are chosen $\omega_{0}/2\pi \approx 4.990\ \mathrm{GHz}$, $\Delta/2\pi\approx 138\ \mathrm{MHz}$, $\nu/2\pi=100\ \mathrm{MHz}$ for all three qubits,
and $\delta\phi \approx 2\pi/3$ between any pair of qubits, which result in zero effective coupling strength
for any pair of qubits, i.e., $H_0 = 0$ in Eq.~(\ref{Heff1}), as confirmed in Fig.~\ref{fig:swap}. With the pulse sequence in Fig.~\ref{fig:expdata_single}(a), we initialize
$|Q_1Q_2Q_3\rangle$ in $|001\rangle$ with a $\pi$ rotation to $Q_3$, then apply the Floquet
modulation signals through qubit Z lines for a period of $T$, and finally tune all
qubits quickly to their idle frequencies for simultaneous readout. As shown in Figs.~\ref{fig:expdata_single}(b) and (c), the measured probabilities for the three single-excitation
multiqubit states $|001\rangle$, $|010\rangle$, and $|100\rangle$ as functions of $T$
are plotted in Fig.~\ref{fig:expdata_single}(b), which indicates that
the excitation circulates along the route $Q_{3} \rightarrow Q_{1} \rightarrow Q_{2} \rightarrow Q_{3}$.
The numerical simulation results taking into account realistic device and experimental
parameters are also plotted for comparison.
In the numerical simulation, the control parameters $\Delta_j$, $\delta\phi_j$, and $\delta\omega_{0,j}$ 
in $Q_j$'s Floquet drive $\omega_{j}(t)=(\bar{\omega}_{0}+\delta\omega_{0,j})+\Delta_j\cos(\nu t+\delta\phi_j)$ are set as 
$\Delta_j/2\pi= \{138\,\textrm{MHz},\, 140\,\textrm{MHz},\, 136\,\textrm{MHz}\}$, 
$\delta\phi_j=\{-0.1,\, 2\pi/3,\, 4\pi/3+0.1\}$, and $\delta\omega_{0,j}/2\pi = \{0,\, 0.7\,\textrm{MHz},\, 0\}$ for $Q_j = \{Q_1,\, Q_2,\, Q_3\}$.
The choice of these parameter values yields a good numerical match with the experimental data, 
which is also reasonable considering how we tune up the experiment.

\begin{figure}[tb]
\centering{}\includegraphics[width=1\linewidth]{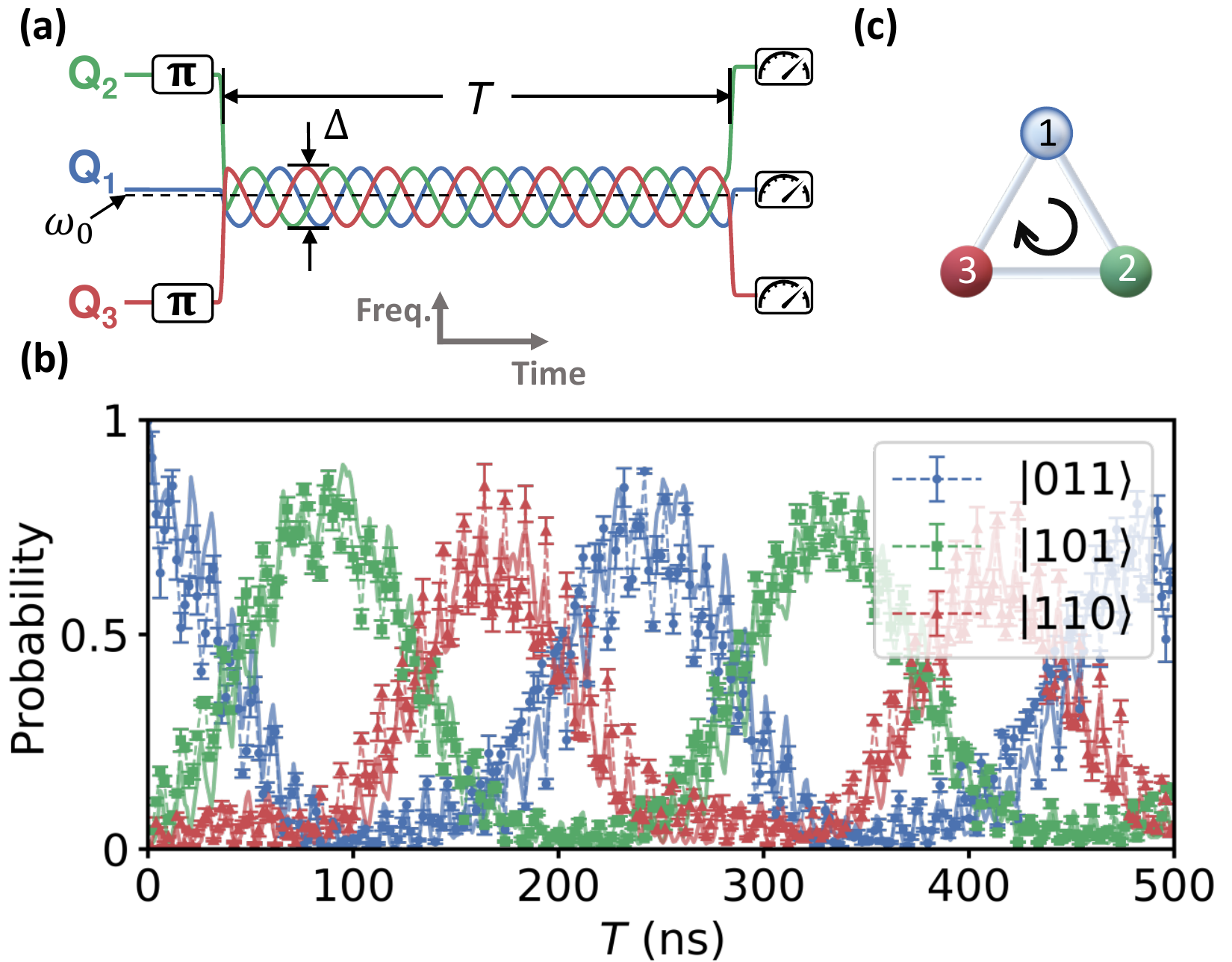} \caption{\label{fig:expdata_two}
(a) Pulse sequence for observing the double-excitation chiral dynamics.
(b) Measured probabilities of the double-excitation multiqubit states (symbols connected by dashed lines)
as functions of the modulation time $T$. Data curves of $P_{011}$, $P_{101}$, and $P_{110}$, 
colored according the colors of the vacancy sites in (c), rise up
sequentially and cyclically, indicating the chiral motion of the vacancy (or the double excitations).
Error bars are standard deviations of five repetitive measurements. Solid lines are the numerical simulation results.
(c) Schematics of the chiral dynamics, where the dark (light) color
represents the qubit in $|1\rangle$ ($|0\rangle$) and the arrow indicates the rotation direction of the vacancy (or the double excitations).}
\end{figure}

For bosons with on-site repulsive interactions in a synthetic gauge
field or spins with DM interactions, the chiral evolution of a double-excitation state, equivalent to the presence of a vacancy in the three-qubit sites,
is opposite to that of a single-excitation state. However, with the
spin chirality interaction, directions of the chiral evolutions of these two states are the
same. To verify this point, here we initialize the three-qubit state in $|011\rangle$ to obtain the double-excitation chiral
dynamics using the pulse sequence shown in Fig.~\ref{fig:expdata_two}(a). We observe that the double excitations collectively
circulate clockwise, i.e., the vacancy in excitations (the $|0\rangle$ state) circulates
along the route $Q_{1} \rightarrow Q_{2} \rightarrow Q_{3} \rightarrow Q_{1}$ as shown in Figs.~\ref{fig:expdata_two}(b) and (c),
in the same direction as that for the single-excitation state in Figs.~\ref{fig:expdata_single}(b) and (c).
However, we notice non-negligible decreases of the modulation amplitudes for all three qubits (here $\Delta/2\pi\approx 135\ \mathrm{MHz}$), 
and meanwhile we observe that the time for completing a full cycle is almost one-half of the time for the single-excitation cycle.
These are due to the existence of the third energy level $|2\rangle$ in the transmon qubits, where the qubit anharmonicity $\eta/2\pi$,
i.e., the frequency difference between the $|1\rangle\leftrightarrow|2\rangle$ and $|0\rangle\leftrightarrow|1\rangle$ transitions, is about ${-230}\ \mathrm{MHz}$.
When there are two excitations, e.g., $|011\rangle$ for $|Q_1Q_2Q_3\rangle$, an additional coupling channel between
$Q_1$ in $|0\rangle$ and $Q_3$ in $|1\rangle$ exists via the second-order effect mediated by $Q_2$'s $|1\rangle\leftrightarrow|2\rangle$,
and the same is between $Q_1$ and $Q_2$. As confirmed by the analysis in Supplementary Material and the numerical
simulation results taking into account the qubit anharmonicity in Figs.~\ref{fig:expdata_two}(b),
the additional second-order effective interaction mediated by $|2\rangle$ indeed
reduces the modulation amplitudes and accelerates the circulation as expected (see Supplementary Material).

In our experiment, the three-body chirality interaction is a second-order effect under the approximation $\nu\gg g$, 
and the resulting interaction strength is estimated to be $\kappa/2\pi\approx 0.5\ \mathrm{MHz}$, 
which is sufficient for the observation of the chiral dynamics. Although this interaction strength is relatively weak and the second-order effect may not allow the implementation of high-fidelity quantum gates, our method of synthesizing the three-body interaction can still be useful for analog quantum simulation, since it is generally difficult to create a dynamical process mimicking that under a many-body Hamiltonian using the experimentally available one- and two-qubit gates (see,
e.g., the recent experiment of generating multicomponent atomic Schr\"odinger cat states of up to 20 superconducting qubits~\cite{Song2019}).

In summary, we experimentally demonstrate a method for synthesizing three-body interaction 
of spin chirality with three superconducting qubits arranged in a triangular loop. 
This method expands the interaction types that we can achieve in superconducting qubits. 
Meanwhile, the models that contain three-body interaction for qubits arranged in triangular
lattices were shown to have multi-degenerate ground states\cite{Pachos2004PRL,Pachos2004PRA,Pachos2008,Lishengwen2010}. 
The degeneracy depends on the topology of the system and is well protected
against perturbations. Our experiment provides one possible way to
realize these models and the ground states of these Hamiltonian can
be used in the implementation of topological physics.\\

\noindent\textbf{Supplementary Material}\\
See Supplementary Material for device parameters, the experiment of tuning the qubit-qubit coupling
by Floquet modulation of just one qubit, the tune-up procedure of the three-qubit Floquet modulations 
for the chiral dynamics, and the detailed theoretical analysis
of the interaction Hamiltonian for both the single- and double-excitation cases.\\

\noindent\textbf{Acknowledgments}\\We thank Hekang Li and Dongning Zheng for
fabricating the device, and Chao Song, Qiujiang Guo, and Zhen Wang for technical support during the experiment.
This work was supported by the National Key Research and Development Program of China (Grants No.~2019YFA0308100)
and the National Natural Science Foundations of China (Grants No. 11725419 and No. 11874322).


\begin{thebibliography}{10}
\bibitem{Feynman1982}R. P. Feynman, ``Simulating physics with computers,'' Int. J. Theor. Phys. \textbf{21},
467 (1982).

\bibitem{Nori2014RMP}I. M. Georgescu, S. Ashhab, and F. Nori, ``Quantum simulation,'' Rev. Mod. Phys. \textbf{86}, 153 (2014).

\bibitem{Somaroo1999}S. Somaroo, C. H. Tseng, T. F. Havel, R. Laflamme, and D. G. Cory, ``Quantum simulations on a quantum computer,'' Phys. Rev. Lett. \textbf{82}, 5381 (1999).


\bibitem{Clarke2008}J. Clarke and F. K. Wilhelm, ``Superconducting quantum bits,'' Nature \textbf{453}, 1031 (2008).


\bibitem{You2011}J. Q. You and F. Nori, ``Atomic physics and quantum optics using superconducting circuits,'' Nature \textbf{474}, 589 (2011).


\bibitem{Devoret2013}M. H. Devoret and R. J. Schoelkopf, ``Superconducting circuits for quantum information: An outlook,'' Science \textbf{339}, 1169 (2013).


\bibitem{Martinis2018Science}C. Neill, P. Roushan, K. Kechedzhi, S. Boixo, S. V. Isakov, V. Smelyanskiy,
A. Megrant, B. Chiaro, A. Dunsworth, K. Arya, R. Barends, B. Burkett,
Y. Chen, Z. Chen, A. Fowler, B. Foxen, M. Giustina, R. Graff, E. Jeffrey,
T. Huang, J. Kelly, P. Klimov, E. Lucero, J. Mutus, M. Neeley, C. Quintana,
D. Sank, A. Vainsencher, J.Wenner, T. C. White, H. Neven, and J. M.
Martinis, ``A blueprint for demonstrating quantum supremacy with superconducting
qubits,'' Science \textbf{360},
195 (2018).


\bibitem{wendin2017}G. Wendin, ``Quantum information processing with
superconducting circuits: A review,'' Rep. Prog. Phys. \textbf{80},
106001 (2017).


\bibitem{VQEprx}P. J. J. O'Malley, R. Babbush, I. D. Kivlichan, J. Romero, J. R. McClean, R.
Barends, J. Kelly, P. Roushan, A. Tranter, N. Ding, B. Campbell, Y. Chen, Z.
Chen, B. Chiaro, A. Dunsworth, A. G. Fowler, E. Jeffrey, E. Lucero, A.
Megrant, J. Y. Mutus, M. Neeley, C. Neill, C. Quintana, D. Sank, A. Vainsencher, J. Wenner, T. C. White, P. V. Coveney, P. J. Love, H. Neven,
A. Aspuru-Guzik, and J. M. Martinis, ``Scalable Quantum Simulation
of Molecular Energies,'' Phys. Rev. X \textbf{6}, 031007 (2016).


\bibitem{Kandala2017}A. Kandala, A. Mezzacapo, K. Temme, M. Takita, M. Brink, J. M. Chow,
and J. M. Gambetta, ``Hardware-efficient variational quantum eigensolver for small molecules and quantum magnets,'' Nature \textbf{549}, 242 (2017).


\bibitem{Kitaev2003}A. Y. Kitaev, ``Fault-tolerant quantum computation
by anyons,'' Ann. Phys. \textbf{303}, 2 (2003).


\bibitem{Goldman2014PRX}N. Goldman, and J. Dalibard, ``Periodically
Driven Quantum Systems: Effective Hamiltonians and Engineered Gauge
Fields,'' Phys. Rev. X \textbf{4}, 031027 (2014).


\bibitem{Martinis2017natphy}P. Roushan, C. Neill, A. Megrant, Y. Chen, R. Babbush, R. Barends, B. Campbell, Z. Chen, B. Chiaro, A. Dunsworth, A. Fowler, E. Jeffrey,
J. Kelly, E. Lucero, J. Mutus, P. J. J. O'Malley, M. Neeley, C. Quintana,
D. Sank, A. Vainsencher, J. Wenner, T. White, E. Kapit, H. Neven, and
J. Martinis, ``Chiral ground-state currents of interacting photons in a synthetic
magnetic field,'' Nat. Phys. \textbf{13}, 146 (2017).


\bibitem{zhuxiaobo2018switch}Y. Wu, L.-P. Yang, M. Gong, Y. Zheng, H. Deng, Z. Yan, Y. Zhao, K. Huang, A. D. Castellano, W. J. Munro, K. Nemoto, D.-N. Zheng, C. P.
Sun, Y. xi Liu, X. Zhu, and L. Lu, ``An efficient and compact switch for quantum circuits,'' npj Quantum Inf. \textbf{4}, 50 (2018).


\bibitem{sunluyan2018statetransfer}X. Li, Y. Ma, J. Han, T. Chen, Y. Xu, W. Cai, H.Wang, Y. Song, Z.-Y. Xue, Z.-q. Yin, and L. Sun, ``Perfect quantum state transfer in a superconducting qubit chain with parametrically tunable couplings,'' Phys. Rev. Appl. \textbf{10}, 054009 (2018).


\bibitem{dawei2019natphy}D.-W. Wang, C. Song, W. Feng, H. Cai, D. Xu, H. Deng, H. Li, D. Zheng, X. Zhu, H. Wang, S.-Y. Zhu, and M. O. Scully, ``Synthesis of antisymmetric
spin exchange interaction and chiral spin clusters in superconducting
circuits,'' Nat. Phys. \textbf{15}, 382 (2019).


\bibitem{SunLuyan2019}W. Cai, J. Han, F. Mei, Y. Xu, Y. Ma, X. Li, H.Wang, Y. P. Song, Z.-Y. Xue, Z.-q. Yin, S. Jia, and L. Sun, ``Observation of topological magnon insulator
states in a superconducting circuit,'' Phys. Rev. Lett. \textbf{123}, 080501 (2019).


\bibitem{Floquet2018}O. Kyriienko and A. S. S{\o}rensen, ``Floquet quantum simulation with superconducting qubits,'' Phys. Rev. Appl. \textbf{9}, 064029 (2018).


\bibitem{Floquet2019}M. Sameti and M. J. Hartmann, ``Floquet engineering
in superconducting circuits: From arbitrary spin-spin interactions
to the Kitaev honeycomb model,'' Phys. Rev. A \textbf{99}, 012333 (2019).


\bibitem{wenxiaogangBOOK}X. G. Wen, \textit{Quantum Field Theory
of Many-body Systems} (Oxford University Press, New York, 2004).


\bibitem{Hafezi2014prb}M. Hafezi, P. Adhikari, and J. M. Taylor, ``Engineering three-body interaction and Pfaffian states in circuit QED systems,'' Phys. Rev. B\textbf{ 90}, 060503 (2014).


\bibitem{wenxiaogang1989}X. G. Wen, F. Wilczek, and A. Zee, ``Chiral
spin states and superconductivity,'' Phys. Rev. B \textbf{39}, 11413
(1989).


\bibitem{Taguchi2001}Y. Taguchi, Y. Oohara, H. Yoshizawa, N. Nagaosa1, and Y. Tokura, ``Spin chirality, berry phase, and anomalous hall effect in a frustrated ferromagnet,'' Science \textbf{291}, 2573 (2001).


\bibitem{Grohol2005}D. Grohol, K. Matan, J.-H. Cho, S.-H. Lee, J. W. Lynn, D. G. Nocera, and Y. S. Lee, ``Spin chirality on a two-dimensional frustrated lattice,'' Nat. Mater. \textbf{4}, 323 (2005).


\bibitem{Bruno2004}P. Bruno, V. K. Dugaev, and M. Taillefumier, ``Topological hall effect
and berry phase in magnetic nanostructures,'' Phys. Rev. Lett. \textbf{93}, 096806 (2004).


\bibitem{Katsura2010}H. Katsura, N. Nagaosa, and P. A. Lee, ``Theory of the thermal hall
effect in quantum magnets,'' Phys. Rev. Lett. \textbf{104}, 066403 (2010).


\bibitem{Hirschberger2015}M. Hirschberger, R. Chisnell, Y. S. Lee, and N. Ong, ``Thermal hall effect of spin excitations in a kagome magnet,'' Phys. Rev. Lett. \textbf{115}, 106603 (2015).


\bibitem{Lyanda2001}Y. Lyanda-Geller, S. H. Chun, M. B. Salamon, P. M. Goldbart, P. D. Han, Y. Tomioka, A. Asamitsu, and Y. Tokura, ``Charge transport in manganites: Hopping conduction, the anomalous hall effect, and universal scaling,'' Phys. Rev. B \textbf{63}, 184426 (2001).


\bibitem{Bauer2014}B. Bauer, L. Cincio, B. Keller, M. Dolfi, G. Vidal, S. Trebst, and A. Ludwig, ``Chiral spin liquid and emergent anyons in a kagome lattice mott insulator,'' Nat. Commun. \textbf{5}, 5137 (2014).


\bibitem{zhou2017}Y. Zhou, K. Kanoda, and T.-K. Ng, ``Quantum spin liquid states,'' Rev.
Mod. Phys. \textbf{89}, 025003 (2017).


\bibitem{Christandl2004}M. Christandl, N. Datta, A. Ekert, and A. J. Landahl, ``Perfect state transfer in quantum spin networks,'' Phys. Rev. Lett. \textbf{92}, 187902 (2004).


\bibitem{Xmon}R. Barends, J. Kelly, A. Megrant, D. Sank, E. Jeffrey, Y. Chen, Y. Yin,
B. Chiaro, J. Mutus, C. Neill, P. O'Malley, P. Roushan, J. Wenner, T. C.
White, A. N. Cleland, and J. M. Martinis, ``Coherent josephson qubit suitable for scalable quantum integrated circuits,'' Phys. Rev. Lett. \textbf{111}, 080502 (2013).


\bibitem{Song2019}C. Song, K. Xu, H. Li, Y.-R. Zhang, X. Zhang, W. Liu, Q. Guo, Z. Wang,
W. Ren, J. Hao, H. Feng, H. Fan, D. Zheng, D.-W. Wang, H. Wang, and
S.-Y. Zhu, ``Generation of multicomponent atomic schrödinger cat states of
up to 20 qubits,'' Science \textbf{365}, 574 (2019).

\bibitem{dawei2016}D.-W. Wang, H. Cai, R.-B. Liu, and M. O. Scully, ``Mesoscopic superposition states generated by synthetic spin-orbit interaction in fock-state
lattices,'' Phys. Rev. Lett. \textbf{116}, 220502 (2016).


\bibitem{Pachos2004PRL}J. K. Pachos and M. B. Plenio, ``Three-spin interactions in optical lattices and criticality in cluster hamiltonians,'' Phys. Rev. Lett. \textbf{93}, 056402 (2004).


\bibitem{Pachos2004PRA}J. K. Pachos and E. Rico, ``Effective three-body interactions in triangular optical lattices,'' Phys. Rev. A \textbf{70},
053620 (2004).


\bibitem{Pachos2008}D. I. Tsomokos, J. J. Garc{\'{\i}}a-Ripoll, N. R. Cooper, and J. K. Pachos, ``Chiral entanglement in triangular lattice models,'' Phys. Rev. A \textbf{77}, 012106 (2008).


\bibitem{Lishengwen2010}Y.-X. Chen, S.-W. Li, and Z. Yin, ``Quantum correlations in a clusterlike system,'' Phys. Rev. A \textbf{82}, 052320 (2010).

\end{thebibliography}
\end{document}

\title{Supplementary Material for ``Synthesizing three-body interaction of spin chirality with superconducting qubits''}
\author{Wuxin Liu}
\thanks{W. L. and W. F. contributed equally to this work.}
\author{Wei Feng}
\thanks{W. L. and W. F. contributed equally to this work.}
\author{Wenhui Ren}
\author{Da-Wei Wang}
\email{dwwang@zju.edu.cn}
\author{H. Wang}
\email{hhwang@zju.edu.cn}
\affiliation{$^{1}$ Interdisciplinary Center for Quantum Information and \mbox{Zhejiang Province Key Laboratory of Quantum Technology and Device,}
\mbox{Department of Physics, Zhejiang University, Hangzhou 310027, China}}

\date{\today}
\maketitle

\section{Device parameters}
Detailed device parameters can be found in Table~\ref{tab:parameters}.
\begin{table}[htb]
    \caption{\label{tab:parameters}
    Device parameters. $\omega_j^\textrm{max}$ is the maximum frequency of $Q_j$. $\omega_j^\textrm{idle}$ is the idle frequency of $Q_j$, where qubit rotations and state readout are performed. $T_{1,j}$ is the energy relaxation time measured for a qubit under the Floquet modulation with $\omega_0/2\pi\approx 4.990$~GHz and $\Delta/2\pi\approx 138$~MHz, and $T_{2,j}^\ast$ is the Ramsey Gaussian dephasing time for a qubit biased at around {4.990}~GHz. Since in the experiment all three qubits are modulated around $\omega_0$ with microwave excitations being transferred among them, we use a much longer pure dephasing time $\sim5$~$\mu$s, instead of $T_{2,j}^\ast$, for all qubits in numerical simulation~\cite{Xu2018}. $\eta_j$ is the qubit anharmonicity. $\omega_j^\textrm{r}$ is the resonant frequency of the readout resonator for each qubit. $F_{0,j}$ ($F_{1,j}$) is the typical probability of measuring $Q_j$ in $|0\rangle$ ($|1\rangle$) when $Q_j$ is prepared in $|0\rangle$ ($|1\rangle$), which is used to correct the measured multiqubit probabilities for elimination of the readout errors.
The $Q_j$-$Q_k$ coupling strength $g_{jk}$ is measured by exciting one of the qubits and then bringing both qubits on resonance at $\approx 4.990$~GHz while all the other qubits are at their maximum frequencies.}
\begin{ruledtabular}
\begin{tabular}{lccc}
    & $Q_1$ & $Q_2$ & $Q_3$  \\
    \hline
    $\omega_j^\textrm{max}/2\pi\ (\mathrm{GHz})$ & 5.767 & 6.113 & 6.062    \\
    $\omega_j^\textrm{idle}/2\pi\ (\mathrm{GHz})$ & 5.025 & 5.380 & 4.650    \\
    {$T_{1,j}\ (\mathrm{\mu s})$} & 18.3 & 12.3 & 16.6    \\
		$T_{2,j}^\ast\ (\mathrm{\mu s})$ & 1.6 & 0.9 & 1.1  \\

    $\eta_j/2\pi\ (\mathrm{MHz})$ & -237 & -234 & -234  \\
    $\omega_j^\textrm{r}/2\pi\ (\mathrm{GHz})$ & 6.541 & 6.659 & 6.687 \\
    $F^0_j$ & 0.984 & 0.932 & 0.957  \\
    $F^1_j$ & 0.939 & 0.906 & 0.938 \\
    \hline
            & $Q_1-Q_2$ & $Q_1-Q_3$ & $Q_2-Q_3$  \\
    $g_{jk}/2\pi\ (\mathrm{MHz})$ & 12.7 & 12.4 & 9.8  \\
    
\end{tabular}
\end{ruledtabular}
\end{table}

\section{single-qubit modulation}
As mentioned in the main text, two qubits can also be dynamically decoupled by modulating only one of them. In this case, the effective coupling strength $g_\mathrm{eff}$ is $gJ_0(\Delta/\nu)$. We confirm this effect by performing an experiment based on the pulse sequence shown in Fig.~\ref{figs1}(a), in which $Q_2$ is modulated with fixed frequency $\nu/2\pi=100\ \mathrm{MHz}$ centering at $Q_1$'s frequency for a period $T$. The probability of state $|01\rangle$ is measured as a function of $T$ with different modulation amplitudes $\Delta$ and the resulting Rabi oscillations between $Q_1$ and $Q_2$ is shown in Fig.~\ref{figs1}(b). In Fig.~\ref{figs1}(c), $|g_\mathrm{eff}|$ as a function of $\Delta$ (dots) is obtained by Fourier transform of the experimental data along the time axis in Fig.~\ref{figs1}(b). When $\Delta/2\pi\approx 240\ \mathrm{MHz}$, the effective coupling strength $|g_\mathrm{eff}|$ becomes zero and the two qubits are dynamically decoupled, which agrees well with the numerical simulation result (line) in Fig.~\ref{figs1}(c). 

\section{procedure of tuning up the three-qubit chiral dynamics}
\begin{figure}[b]
	\includegraphics[width=\linewidth]{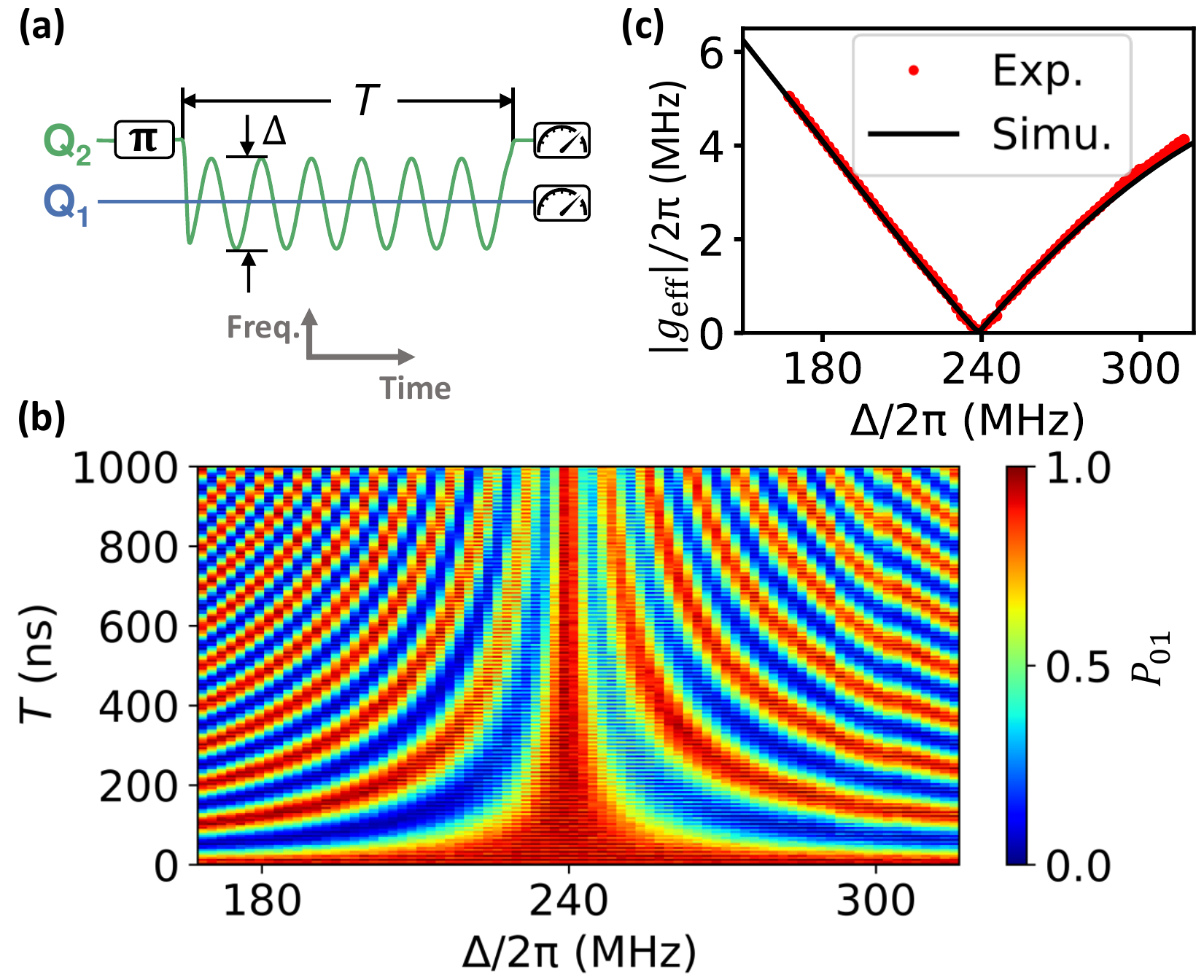}
	\caption{(a) Pulse sequence with the single-qubit modulation to tune the effective coupling strength $g_\mathrm{eff}$ between $Q_1$ and $Q_2$. Both qubits are initialized in $|0\rangle$ while all the other qubits on the same chip are far detuned. Then $Q_2$ is excited to $|1\rangle$ by a $\pi$ rotation and after that its frequency is sinusoidally modulated for a period $T$, during which $Q_1$'s frequency is fixed at $\approx 4.990 \ \mathrm{GHz}$. The center of the sinusoidal modulation is fixed at $Q_1$'s frequency and the modulation frequency $\nu/2\pi=100\ \mathrm{MHz}$ is also fixed while the modulation amplitude $\Delta$ is varied. Finally the probability of the $Q_1$-$Q_2$ joint state $|01\rangle$, $P_{01}$, is measured. (b) $P_{01}$ measured as a function of $T$ and $\Delta$. (c) Effective coupling strength $|g_\mathrm{eff}|$ as a function of $\Delta$ (dots), obtained by Fourier transform of the data in (b), in comparison with the numerical simulation result (line).}
	\label{figs1}
\end{figure}

The procedure to obtain the chiral dynamics (Figs. 3 and 4 of the main text) consists of three steps as sequentially illustrated in Fig. 1, Fig.~\ref{figs1}, and Fig. 2.

\emph{Step 1}: We measure the resonant frequency $\omega_j$ versus Z bias curve for each qubit (Fig. 1 of the main text). \emph{Step 2}: For a pair of qubits, we modulate one qubit in frequency centering around the other one which is fixed in frequency (Fig.~\ref{figs1}), according to the $\omega_j$ versus Z bias curves obtained in step 1. The measurement results can be used to calibrate the drive amplitudes $\Delta$, since the decoupling amplitude is supposed to be 2.404 times the modulation frequency. \emph{Step 3}: With the drive amplitudes being calibrated for the pair of qubits, we then modulate both qubits simultaneously while varying the relative phase of the two modulation drives (Fig. 2 of the main text). The measurement results can be used to calibrate the relative phases since the decoupling point is supposed to be $2\pi/3$. 

Using the drive amplitudes and relative phases calibrated above for pairs of qubits, we can assemble the experimental pulse sequences and measure the 3-qubit chiral dynamics data, which clearly demonstrate chiral flows of the qubit excitations as expected. We then fine-tune the drive amplitude, phase, and offset for each qubit around the calibrated values to maximize the oscillation amplitudes, i.e., for the best quality of the data as shown in Figs. 3 and 4 of the main text. In the numerical simulation we also fine-tune the amplitude, phase, and offset of the Floquet modulation around the nominal experimental values for each qubit to obtain a good numerical match with the experimental data.

\section{Derivation of the effective hamiltonian}

When the frequencies of the three qubits are modulated according to
$\omega_{j}(t)=\omega_{0}+\Delta\cos(\nu t+2\pi j/3)$, the Hamiltonian
in the interaction picture is
\begin{eqnarray}
H_{I} & = & \sum_{jk}\hbar g\sigma_{j}^{+}\sigma_{k}^{-}e^{if\left[\sin\left(\nu t+2\pi j/3\right)-\sin\left(\nu t+2\pi k/3\right)\right]}+\textrm{h.c.}\nonumber \\
 & = & \sum_{jk}\hbar g\sigma_{j}^{+}\sigma_{k}^{-}e^{if2\sin\left(\pi/3\right)\cos\left[\nu t+\frac{\left(j+k\right)\pi}{3}\right]}+\textrm{h.c.}
\end{eqnarray}
where $f=\Delta/\nu$ . The above Hamiltonian can be expanded to $H_{I}=\sum_{n}H_{n}e^{in\nu t}$
with
\begin{equation}
H_{0}=\hbar gJ_{0}\left(\sqrt{3}f\right)\sum_{jk}\sigma_{j}^{+}\sigma_{k}^{-}+\textrm{h.c.},
\end{equation}

\begin{equation}
H_{n}=\hbar gi^{n}J_{n}\left(\sqrt{3}f\right)e^{\frac{in(j+k)\pi}{3}}\sum_{jk}\left[\sigma_{j}^{+}\sigma_{k}^{-}+\left(-1\right)^{n}\sigma_{j}^{-}\sigma_{k}^{+}\right].
\end{equation}
Under the condition $\nu\gg g$, we obtain the effective Floquet Hamiltonian
\cite{PRX2014}

\begin{align}
H_{\textrm{eff}} & \approx H_{0}+\sum_{n=1}^{\infty}\frac{1}{n\hbar\nu}\left[H_{n},H_{-n}\right],
\end{align}
where

\begin{align}
\left[H_{n},H_{-n}\right]= & \sum_{jkl}\hbar^{2}g^{2}J_{n}^{2}\left(\sqrt{3}f\right)\sin\left(\frac{n\pi}{3}\right)\sigma_{j}^{z}\left(\sigma_{k}^{x}\sigma_{l}^{y}-\sigma_{k}^{y}\sigma_{l}^{x}\right).
\end{align}
By setting $J_{0}\left(\sqrt{3}f\right)=0$ with $\Delta/\nu=2.404/\sqrt{3}$,
we obtain $H_{0}=0$, and
\begin{equation}
H_{\textrm{eff}}=\hbar\kappa\hat{\chi},\label{eq:9}
\end{equation}
where $\kappa=g^{2}\beta/\nu$ and $\beta=\sum_{n=1}^{\infty}J_{n}^{2}(\sqrt{3}f)\sin(n\pi/3)/n$.

\section{Effect of weak anharmonicity}

The transmon qubit is not a pure two-level system, which has multiple energy levels and 
the energy separations between adjacent two levels are not equal. 
Qubit anharmonicity is defined by $\eta=\omega^{21}-\omega^{10}$, where $\omega^{lm}$ is 
the transition frequency between levels $l$ and $m$ ($l,m=0,1,2,...$, and the ground state corresponds to level 0).
Here we consider the effect of $\eta$ by including three energy levels, 
$\left\{ \left|0\right\rangle, \left|1\right\rangle, \left|2\right\rangle \right \}$, for each transmon qubit.
Note that when the sinusoidal modulation pulse is applied to the transmon qubit,
both $\omega^{21}$ and $\omega^{10}$ vary simultaneously over time and $\eta$ remains almost a constant.
%along with the transition frequency $\omega_{j}\left(t\right)=\omega_{j}^{10}\left(t\right)$
%being modulated, the transition frequency $\omega_{j}^{21}\left(t\right)$
%is also modulated to make the relationship $\omega_{j}^{21}\left(t\right)-\omega_{j}^{10}\left(t\right)=\eta$
%fixed. 
The full Hamiltonian is
\begin{equation}
\begin{aligned}
H & =  \hbar\sum_{j=1}^{3}\left[\omega_{j}^{10}(t)\left|1\right\rangle _{j}\left\langle 1\right|_{j}+\left[2\omega_{j}^{10}(t)+\eta_j\right]\left|2\right\rangle _{j}\left\langle 2\right|_{j}\right]\\
&+\hbar g\sum_{jk}\left(\left|1\right\rangle _{j}\left\langle 0\right|_{j}\left|0\right\rangle _{k}\left\langle 1\right|_{k}+\textrm{h.c.}\right) \\
 & +\sqrt{2}\hbar g\sum_{jk}\left(\left|2\right\rangle _{j}\left\langle 1\right|_{j}\left|0\right\rangle _{k}\left\langle 1\right|_{k}+\textrm{h.c.}\right)\\
 &+2\hbar g\sum_{jk}\left(\left|2\right\rangle _{j}\left\langle 1\right|_{j}\left|1\right\rangle _{k}\left\langle 2\right|_{k}+\textrm{h.c.}\right),
\end{aligned}
\end{equation}
which is used in numerical simulation.

When there is no population in $\left|2\right\rangle$ of any of the three qubits during the qubit initialization stage, 
the total effective Hamiltonian of the system under the Floquet modulation pulses
can be restricted to the subspace $\left\{ \left|0\right\rangle ,\left|1\right\rangle \right\}$ for each qubit, and under the assumption $|n\nu+\eta|\gg gJ_n(\sqrt 3f)$ we have
\begin{equation}
H_{\mathrm{eff}}\approx H_{0}+\hbar\kappa\hat{\chi}+\left[\hbar\kappa^{\prime}\left(S_{z}+1/2\right)\sum_{jk}\sigma_{j}^{+}\sigma_{k}^{-}+\textrm{h.c.}\right],
\end{equation}
where $S_{z}=\sum_{j}\sigma_{j}^{z}/2$, $\kappa^{\prime}=-2g^{2}\sum_{n=-\infty}^{+\infty}\frac{J_{n}^{2}(\sqrt{3}f)}{n\nu+\eta}e^{in\pi/3}$
is an additional second-order coupling between qubits due to the existence of $\left|2\right\rangle$. 
We note that, for $n=2$, $|n\nu+\eta| \approx 6 gJ_n(\sqrt 3f)$ in our experiment and the assumption may not be well satisfied.
However, our numerical simulation results are based on the full Hamiltonian of Eq. (1) in the main text without any approximation, and the apparent agreement 
with the experimental data indicates that our theory still works decently.

In general, $\kappa^{\prime}$
is a complex number and it can be rewritten as $\kappa^{\prime}=\alpha+i\lambda$.
With $\hat{E}_{s}=\sum_{jk}\left(\sigma_{j}^{+}\sigma_{k}^{-}+\sigma_{j}^{-}\sigma_{k}^{+}\right)$
being the symmetric exchange interaction and $\hat{E}_{as}=i\sum_{jk}\left(\sigma_{j}^{+}\sigma_{k}^{-}-\sigma_{j}^{-}\sigma_{k}^{+}\right)$
being the antisymmetric exchange interaction, the effective Hamiltonian
can be written as (note that $\hat{E}_{as}=S_{z}\hat{\chi}$)
\begin{align}
H_{\mathrm{eff}}= & H_{0}+\hbar\kappa\hat{\chi}+\frac{1}{2}\hbar\lambda\left(S_{z}\hat{\chi}+\frac{1}{2}\hat{\chi}\right)+\hbar\alpha\left(S_{z}+1/2\right)\hat{E}_{s}\nonumber \\
=H_{0}+ & \hbar\left[\left(\kappa+\frac{\lambda}{4}\right)+\frac{\lambda}{2}S_{z}\right]\hat{\chi}+\hbar\alpha\left(S_{z}+1/2\right)\hat{E}_{s}.
\end{align}
Using the experimental parameters, $\nu=100\thinspace\mathrm{MHz}$, $\Delta=138\thinspace\mathrm{MHz}$, and
$\eta=-234\thinspace\mathrm{MHz}$, we have $J_{0}\left(\sqrt{3}f\right)\approx0$, $\alpha\approx0$, and
\begin{equation}
H_{\mathrm{eff}}\approx\hbar\left[\left(\kappa+\frac{\lambda}{4}\right)+\frac{\lambda}{2}S_{z}\right]\hat{\chi}.
\end{equation}

Here $\alpha$ is actually slightly negative. Since $H_{0}$ contains the $\hat{E}_{s}$ term 
with the coefficient $gJ_{0}\left(\sqrt{3}f\right)$, to cancel this negative $\alpha$ 
we can reduce the modulation amplitudes so that $gJ_{0}\left(\sqrt{3}f\right)$ becomes slightly positive, as experimentally observed. In the numerical simulation results in Fig. 4 of the main text, we choose $\Delta_j/2\pi= \{135\,\textrm{MHz},\, 137\,\textrm{MHz},\, 133\,\textrm{MHz}\}$ for $Q_j = \{Q_1,\, Q_2,\, Q_3\}$ while all other parameters remain the same as those used in the numerical simulation of Fig. 3, and a good agreement between the numerical results and the experimental data is obtained.

In the case of a single excitation, $S_{z}=-1/2$ and the coupling
strength is $\kappa$. However, the coupling strength is $\kappa+\lambda/2$
with $S_{z}=1/2$ when there are two excitations. Therefore,
the circulation for the two-excitation case is faster than that for the single-excitation case with $\lambda>0$, 
as experimentally observed in Figs. 3 and 4 of the main text.